# Abrupt changes of hydrothermal activity in a lava dome detected by combined seismic and muon monitoring[1]

Y. Le Gonidec[1], M. Rosas-Carbajal[2], J. de Bremond d'Ars[1], B. Carlus[3], J.-C. Ianigro[3], B. Kergosien[1], J. Marteau[3] & D. Gibert[1,4,*]

[1] Univ Rennes, CNRS, Géosciences Rennes - UMR 6118, F-35000 Rennes, France. [2] Institut de Physique du Globe de Paris, CNRS - UMR 7154, F-75005 Paris, France. [3] Univ Claude Bernard, CNRS, Institut de Physique Nucléaire de Lyon - UMR 5822, Lyon, France. [4] National Volcano Observatory Service, CNRS, OSUR - UMS 3343, F-35000 Rennes, France.

**The recent 2014 eruption of the Ontake volcano in Japan recalled that hydrothermal fields of moderately active volcanoes have an unpredictable and hazardous behavior that may endanger human beings. Steam blasts can expel devastating ejecta and create craters of several tens of meters. The management of such hydrothermal hazards in populated areas is problematic because of their very short time constants. At present no precursory signal is clearly identified as a potential warning of imminent danger. Here we show how the combination of seismic noise monitoring and muon density tomography allows to detect, with an unprecedented space and time resolution, the increase of activity of a hydrothermal focus located 50 to 100 m below the summit of an active volcano, the La Soufrière of Guadeloupe, in the Lesser Antilles. The present study deals with hydrothermal activity events at timescales of few hours to few days. We show how the combination of those two methods improves the risk evaluation of short-term hazards and the localization of the involved volumes in the volcano. We anticipate that the deployment of networks of various sensors including temperature probes, seismic antennas and cosmic muon telescopes around such volcanoes could valuably contribute to early warning decisions.**

Phreatic and hydrothermal eruptions frequently occur in volcanic geothermal fields[1]. This type of events is driven by the rapid expansion of water flashing to steam due to either overheating or decompression, while magma remains at depth. The destabilization of the hydrothermal system may lead to a sudden decompression and eventually evolve into a non-magmatic explosive eruption driven by hydrothermal fluids, liquid or vapor[2]. Although most volcanic hazard studies concern magmatic eruptions, events involving no magma emission gain a growing attention because of the occurrence of recent eruptions[1], such as the laterally-directed explosions that caused at least 58 fatalities at Ontake volcano (Japan) in 2014[3]. A similar eruption occurred at the Te Maari Crater in Mount Tongariro (New Zealand) in 2012[4]. These catastrophic events recalled that ejecta produced by hydrothermal blasts may be as fast as magmatic ones and that phreatic eruptions represent dangerous hazards, difficult to forecast[3]. It is therefore important to identify precursors of unrest in volcanic shallow hydrothermal systems[5], with the difficulty to derive generalized concepts because of the specificity of each volcanic system[6]. Volcanoes with moderate activity and well-developed hydrothermal systems often experience decades of intense hydrothermal activity smearing out potential warning signals announcing an imminent destabilization[7,8]. The detection of such transient signals, possibly very localized in time and space, together with the quantification of hydrothermal volumes and amount of energy involved, constitute new challenges for modern volcanology. The detection of early warning signals requires to monitor the behavior of volcanic systems on short time scales of less than one hour to identify a situation of imminent destabilization. The complexity of the involved mechanisms and the high heterogeneity of volcanic lava domes prevent from using purely deterministic approaches based, for instance, on mechanical models and failure thresholds.

---

[1] Submitted to Nature Scientific Reports, November 2018.

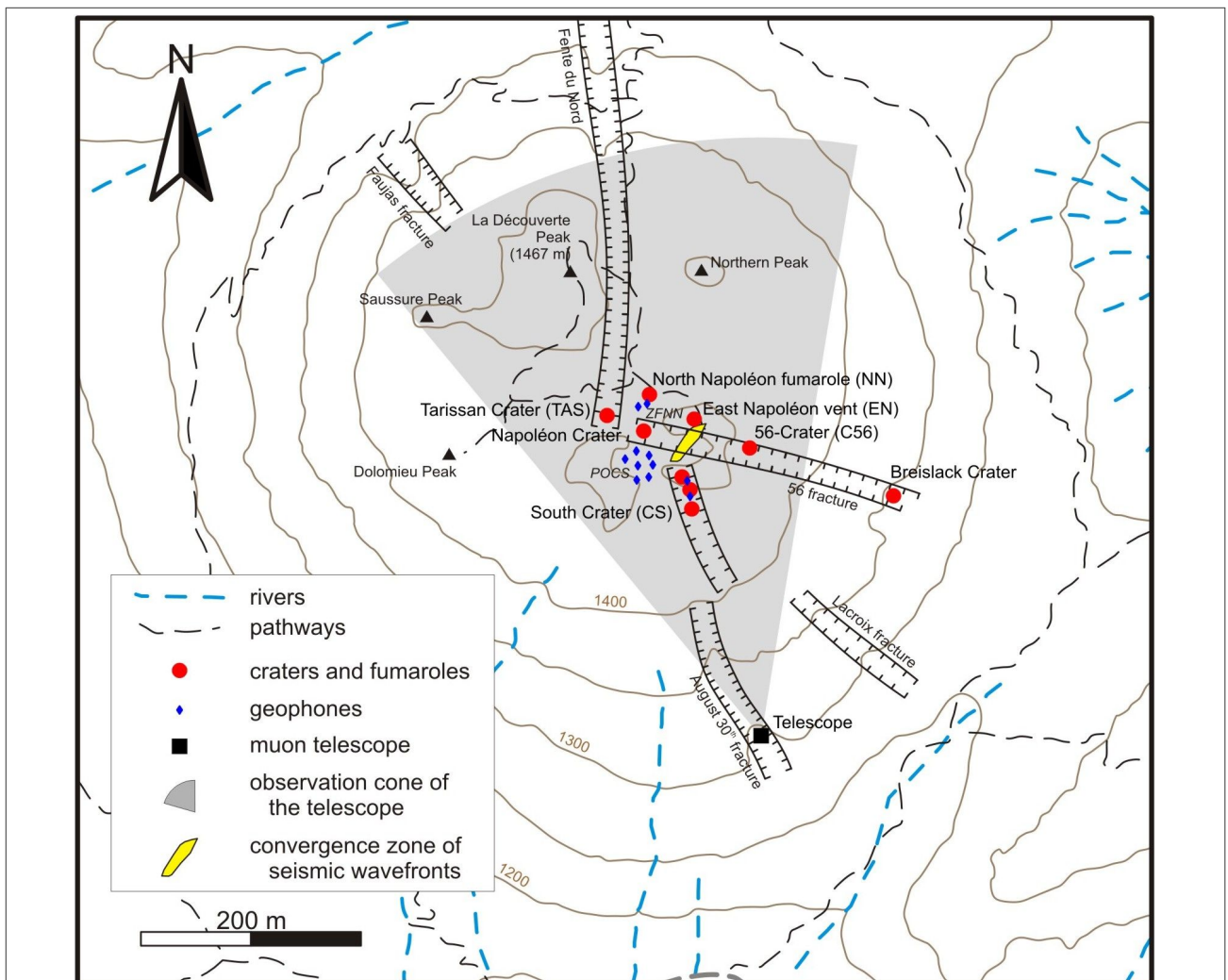

**Fig. 1** Main structures of the La Soufrière lava dome with sensors emplacements. The gray curved triangle represents the view field of the muon telescope (black square) placed in the 30 August fracture at the apex of the triangle. The vents are shown as red dots, the geophones as blue diamonds, and the green patch is the plan view of the source of seismic noise (see Fig. 3).

We conceived an experiment on the La Soufrière volcano in Guadeloupe to answer the questions raised above and, more specifically, those concerning the detection of very short-term warning signals (i.e. time scales of hours and days) of possible hydrothermal destabilization in an active lava dome of moderate activity. La Soufrière belongs to the Lesser Antilles volcanic arc, and the present lava dome was created about 500 years ago, following eight dome collapse events that occurred during the last 8500 years[9,10]. Several collapses produced laterally-directed explosions caused by blasts of hydrothermal fluids expanding laterally at estimated speeds of 100-230 m/s. The last 1976-1977 eruption is considered as a failed magmatic event where a small andesitic magma volume stopped its ascent approximately 3 km below the surface[11,12]. Degassing then decreased to reach a minimum in 1991 before increasing again with an intense fumarolic activity at the summit in 1992 and then, a sudden onset of chlorine degassing from the South Crater vents in 1998 (Fig. 1). In 2014, a new active region appeared to the East of the Tarissan crater and the activity of the Napoléon crater and the 56-crater increased (Fig. 1). This recent evolution may be due to rearrangement of flow paths in response to the progressive hydrothermal sealing of open fractures causing over-pressurization until new paths are opened by hydro-fracturing[13]. The recent hydrothermal activity at the summit of the lava dome is compatible with an increasingly vapor-dominated system favorable to local destabilization as shown by at least two small explosions that occurred in 2016 in the East-Napoléon vent (Fig. 1). Such hazardous environments result from rapid phenomena occurring inside the lava dome at time scales of few hours or days.

# Results

**Vent temperature and seismic noise data**.

To detect potential early warning signals, we use seismic and temperature data acquired with our high time-resolution (4 ms) arrays of sensors located on the summit of La Soufrière lava dome together with one of our 6 cosmic muon telescopes, located inside the 30-August fracture (Fig. 1). Because the hydrothermal activity changes without clear warning signals, permanent recordings are required . This unique data set focuses on a particular period in 2017, from March 28[th] 12:00 UTC to March 31[st] 12:00 UTC, characterized by a remarkable oscillation sequence of some seismic attributes in correlation with delayed time-variations of fumarole temperature. This original multi-sensor correlation is the first identification of a short-term transient signal of the La Soufrière volcano presently with a moderate fumarolic and non-explosive activity . We focus our analysis on this 3-day period by combining temperature, seismic and muon monitoring to relate the transient signature of the hydrothermal activity to physical properties. We observe a moderate increase of the thermal regime in the South crater vents (Fig. 1), particularly visible in the measured temperature in one of those vents rising from 98.1°C to 99.3°C through a sequence of oscillations of increasing period (Fig. 2A). The two other vents of this crater display the same pattern, which is also visible in the time variations of the seismic noise root mean square (RMS) (Fig. 2B). During the considered 3-day period, the seismic data do not suffer from perturbations due to either wind or rain.

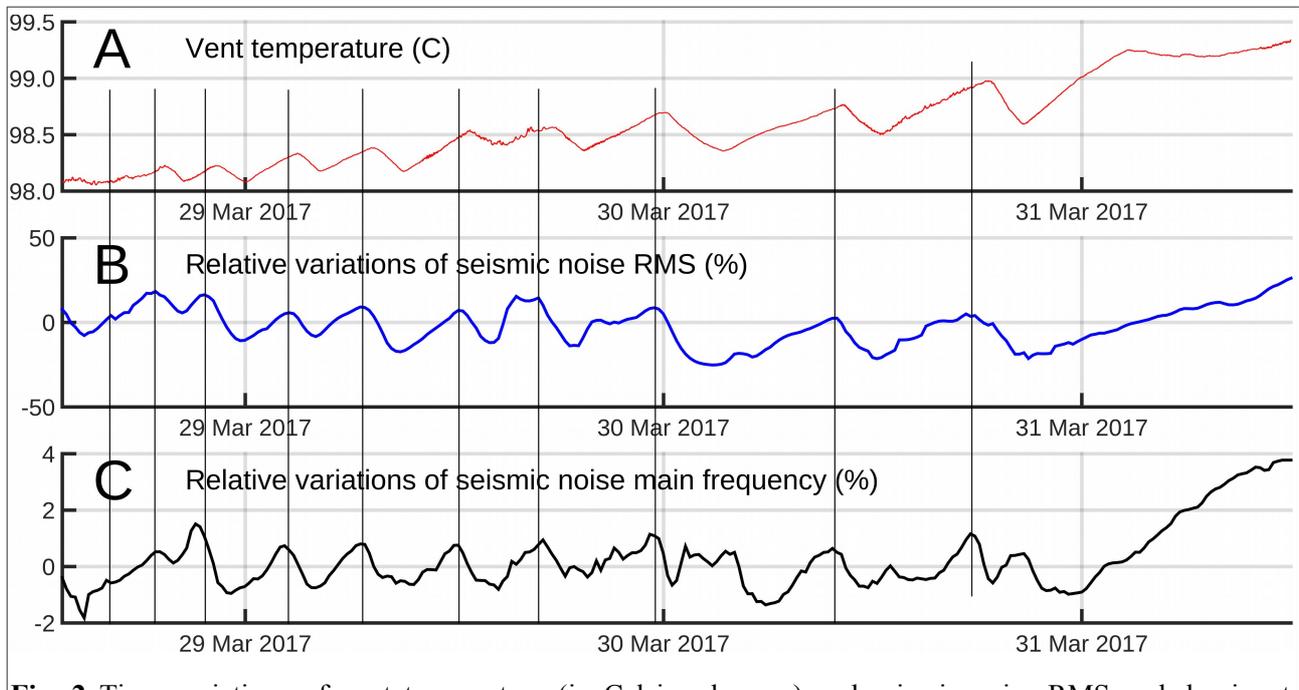

**Fig. 2** Time variations of vent temperature (in Celsius degrees) and seismic noise RMS and dominant frequency. **a** Temperature in the North vent of the South crater. **b** Relative variations (in percent) of the seismic noise energy in the 3-6 Hz frequency band. **c** Relative variation (in percent) of the dominant frequency in the 3-6 Hz spectral band. The thin black vertical lines mark the relative maxima of seismic RMS that fall nearby temperature maxima. Observe the delay of about 45 mn of the temperature maxima with respect to those of the seismic noise RMS. The sampling interval of temperature data is 1s and the seismic attributes (RMS and dominant frequency) are computed for time windows of 20s.

The same oscillating pattern is also visible in the dominant frequency of the seismic noise corrected for the gain function of the vertical geophones (Fig. 2C). It can be observed that the oscillations of the seismic noise RMS are in advance of 43±11 mn with respect to the temperature oscillations. A similar advance of 48±14 mn is found for the noise dominant frequency curve. Assuming that the vent temperature is correlated with the steam pressure, we infer that, if produced by the vent itself, the seismic noise should be in phase with temperature. Consequently, the observed time shift eliminates the possibility that the vents themselves may be the source of the seismic noise. To localize the source of the seismic noise, a spectral coherency analysis[14] is performed, followed by a bandpass filtering to isolate the most correlated waveband in the range 3 Hz < f < 25 Hz and determine the time lags among the signals recorded with the array of geophones. The time lags are then used in a back-propagation modelling that does not constrain the solution: non-physical solutions located outside the lava dome, i.e. in the atmosphere, could be obtained in case of incoherent time lags. Using this procedure, we find that the refocusing of the wave-fronts inside the lava dome is achieved only for seismic wave velocities around 500 m/s and occurs in a rather small volume of about $10^4$ m$^3$ (Fig. 3).

**Cosmic muons radiography.**

The source of the seismic noise associated to the oscillation sequence (Fig. 2) is located inside the observation cone of the cosmic muon telescope located inside the 30 August fracture (Fig. 1) and dedicated to high-resolution imaging of the hydrothermal system located beneath the South, Tarissan and Gouffre 56 craters. Let us recall that this instrument counts the cosmic muons crossing the volcano along a discrete set of lines of sight[15,16]. The incident muon flux is attenuated according to the volcano's opacity in $kg \cdot m^{-2}$ [17]. In practice, the larger the number of muons, the lower the opacity and the average density. The telescope used in the present analysis is placed at an altitude of 1268 m, i.e. about 150 m below the South crater area, with an inclination of 28.5° and azimuth of 345°. It covers a solid angle encompassing the South crater, the Tarissan crater and the 56-crater (Fig. 1). The telescope counts 3 matrices with $10 \times 10$ pixels of $5 \times 5$ cm$^2$ that allow to simultaneously scan $19 \times 19 = 361$ lines of sight covering the instrument's viewing solid angle.

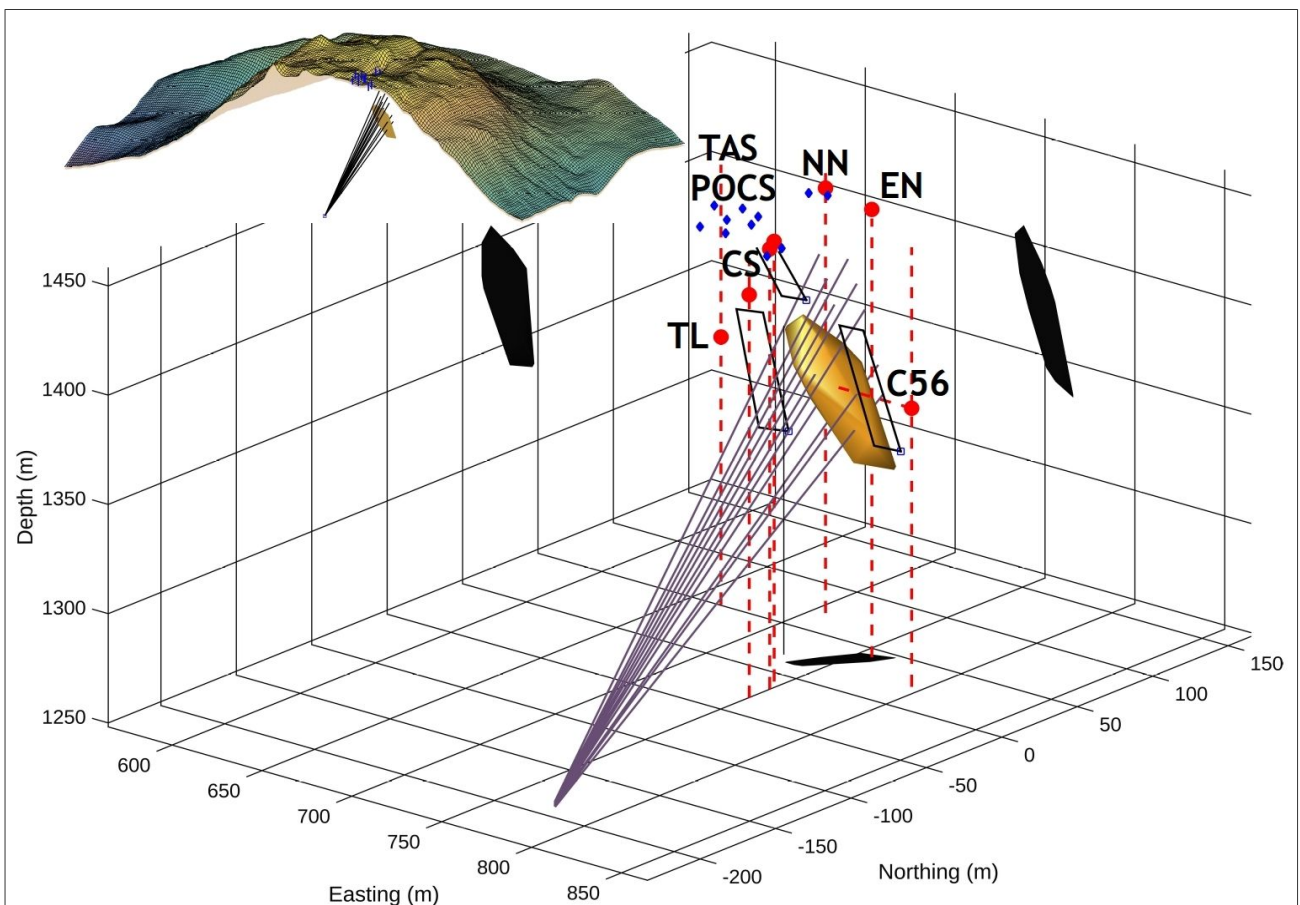

**Fig. 3** Location of the seismic noise source volume. The yellow body represents the 3D convergence zone of the seismic wavefronts recorded by the geophones of the POCS, CS and NN areas on top of the lava dome (blue dots). This volume is reconstructed by back-propagating the time lags of seismic signals and is likely to contain the sources of the seismic noise. The red dots represent the main active vents. The black patches are the projections of the source volume onto the faces of the 3D block diagram. The fan-like bundle of straight lines represents the lines of sight of the muon telescope crossing the active hydrothermal

region and used to obtain the red curve in Fig. 4. The 3 black rectangles located above and on each sides of the source zone show the 3 adjacent areas corresponding to the curves labeled 2, 3 and 4 in Fig. 4. The telescope is located at the apex of the fan-like pattern. The inset shows the position of the seismic source zone in the lava dome (see Extended Data Fig. 6 for an enlarged version). The red dots mark the main fumaroles TAS = Tarissan crater, TL Tarissan acid pond, C56 = 56-crater fumaroles, CS = South crater fumaroles, NN = North Napoléon fumarole, EN = East Napoléon fumarole. POCS = main array of geophones used in the present study. The vertical dashed red lines passing through the fumaroles markers are upward-continued to the top of the lava dome to better show that TL and G56 are located in pits about 80 m below the surface.

The resolution of opacity variations is related to the time resolution through a feasibility formula including also the total opacity and the telescope's acceptance, which scales roughly like the detection matrices section[18]. In the present study, the time-resolution is about 10-20 days and the variance of the muons counts can be improved by merging adjacent lines of sight in order to increase the acceptance[19]. Consequently, the muons count time-series goes from January 8$^{th}$ to April 14$^{th}$ 2017 in order to largely encompass the 3-day period where we evidenced the correlated events in the temperature and seismic data (Fig. 2). Fig. 4 displays the muons count time series in 4 areas: one centered on the source of the seismic noise (red curve of Fig. 4), and three nearby areas located on the left, on the right and above (blue curves in Fig. 4). It can be observed that a conspicuous increase of the muons count occurs in the region of the seismic source while the other three areas remain stationary. This sharp increase of muons count indicates a dramatic decrease of opacity starting at the very beginning of April 2017. Because of the averaging performed to improve the statistics of the muons count, the time-resolution of the curves displayed in Fig. 4 are not as fine as those in Fig. 2. However, we remark that the sharp increase of flux occurs on April 4$^{th}$ and that a possible moderate increase may occur as soon as March 31$^{st}$ when the descending phase of the flux oscillation is interrupted. Consequently, there is possibly a delay of at least two days between the beginning of the temperature and seismic oscillations shown in Fig. 2 and the onset on increase of muon flux. The stationary fluctuations of the muon flux observed in the four domains before the sharp increase in domain 1 are likely to correspond to density variations caused by interactions between supplies of meteoric water and regular hydrothermal activity in the lava dome.

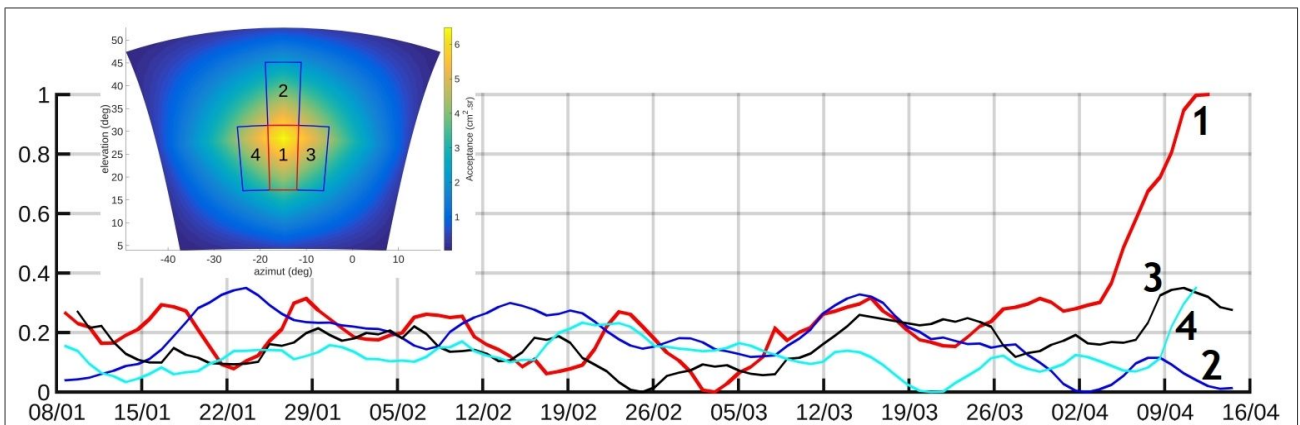

**Fig. 4** Time variations of the muons flux across different domains of the lava dome. The red curve is for the bundle of lines of sight covering the seismic source zone of Fig. 3. The other three curves are for adjacent areas labeled 2, 3 and 4 in the inset showing the acceptance function of the telescope's view-field. Oscillation amplitudes are arbitrarily set to a common value. The total acceptance of bundles of lines of sight crossing the source area 1 is $54.3\,cm^2\,sr$. The acceptances of the merged lines of sight in the areas 2, 3 and 4 are respectively $34.8\,cm^2\,sr$, $40.3\,cm^2\,sr$, and $45.8\,cm^2\,sr$. For comparison, the maximum acceptance of the axial line of sight of the telescope equals $6.5\,cm^2\,sr$.

**Dynamics of the shallow hydrothermal system**.

The remarkable oscillation pattern observed in the time series of both temperature and seismic noise attributes (Fig. 2) suggests a causal link between the temperature measured in the vents and the activity of the source of seismic noise localized using both the seismic and the muons data (Fig. 3 and 4). This source zone appears vertically elongated over more than 50 m and slightly inclined Southward. Such a shape is partly a consequence of the geometry of the geophone array which is located only on the Western side of the source zone. Its lower part is connected to the fracture 56 which extends from the Napoléon crater to the West to the Breislack crater to the East (Fig. 1 and 5a). The sharp decrease in opacity observed in the source zone located within the lava dome may be explained by the rapid invasion of steam flushing a liquid phase (Fig. 5b). This is likely due to the convective destabilization of hydrothermal fluids triggering the ascend of hot fluids with an increasing gas fraction in the deep part of the fracture 56 network. These hot fluids are likely to reside in the lava dome and in the first hundreths of meters below. In this scenario, the oscillations shown on Fig. 2 are early-warning phenomena that occur at the very beginning of the process with the first signs of destabilization in the source zone (Fig. 5a). We check this idea by analyzing the temperature oscillations as a log-periodic sequence[20] to estimate the occurrence time of an eventual singular event marking the destabilization. We find this occurrence about 15 h before the appearance of the first visible oscillation.

The time-variations of the dominant frequency of the seismic noise (Fig. 2C) may reflect long-period changes of the physical conditions in a resonator formed by a fracture network filled with a liquid-bubble mixture of high bulk compressibility, high density and low sound velocity (Fig. 5). The small variations of the dominant frequency may be reproduced by varying the sound velocity in a Helmholtz-like resonator system. Using the data modeling results shown if the Fig. 9b of the article published by Kieffer[21], we derive the mass fraction $\varphi$ of steam necessary to reproduce the observed frequency variations. We find that a minimum mass fraction $\varphi = 11\%$ is necessary to obtain a physical solution and that variations of 0.25% above this value are sufficient to reproduce the data. The frequency jump visible after March 31$^{st}$ (Fig. 2C) requires a $\varphi$ increase of 0.75%. Such a mass fraction corresponds to a volume fraction of steam around 97% at pressures less than 10 bars. This would imply a steam flow in either a slug or annular flow regime[22]. These types of flow are known for their violent dynamics[23] and for their efficient seismic emissivity[24]. Two-phase flows in conduit networks are reported to be intrinsically unstable and prone to oscillations sustained by turbulent movements and density waves[25,26]. In many instances, the period of the density waves is proportional to the wave time-of-flight required to cross the system[27]. Within this phenomenological context, the observed increase of the oscillation period in the temperature and seismic times series (Fig. 2) may reflect a progressive extension of the size of the fracture network occupied by the two-phase flow (Fig. 5a). This extension may result from an enhanced extraction of the gaseous phase caused by the pressure oscillations[28] that may produce local negative pressure anomalies in the deep parts of the draining network. This kind of positive feedback could eventually lead to an abrupt jump of steam production exceeding the transport capacity of the surface vents and leading to explosive events. Owing to these observations, we consider that the fracture 56 of La Soufrière lava dome is prone to destabilization and represents an area of potentially high risk level. The temperature trend observed along the cycle of temperature oscillations (Extended Data Fig. 2) may be due to a constant steam supply causing a progressive increase of the pressure in the source zone of Fig. 3.

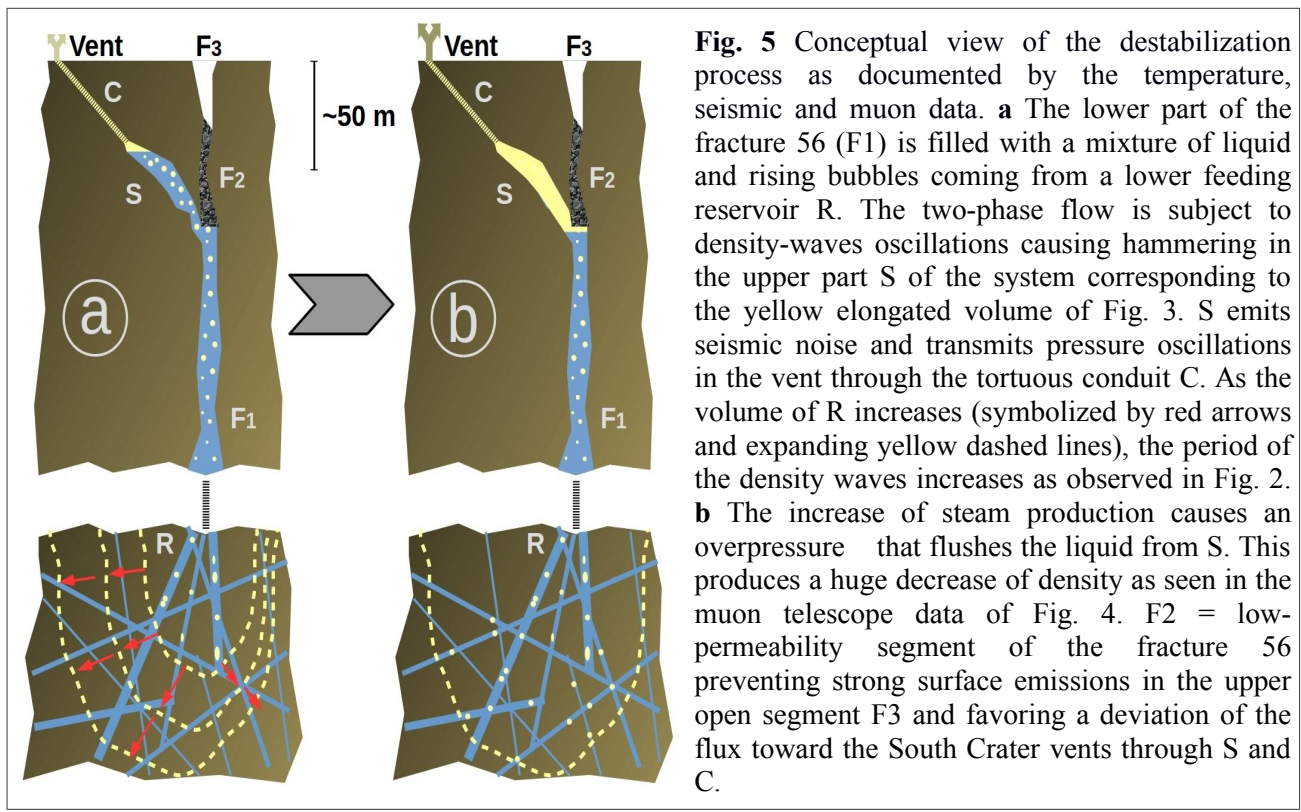

**Fig. 5** Conceptual view of the destabilization process as documented by the temperature, seismic and muon data. **a** The lower part of the fracture 56 (F1) is filled with a mixture of liquid and rising bubbles coming from a lower feeding reservoir R. The two-phase flow is subject to density-waves oscillations causing hammering in the upper part S of the system corresponding to the yellow elongated volume of Fig. 3. S emits seismic noise and transmits pressure oscillations in the vent through the tortuous conduit C. As the volume of R increases (symbolized by red arrows and expanding yellow dashed lines), the period of the density waves increases as observed in Fig. 2. **b** The increase of steam production causes an overpressure that flushes the liquid from S. This produces a huge decrease of density as seen in the muon telescope data of Fig. 4. F2 = low-permeability segment of the fracture 56 preventing strong surface emissions in the upper open segment F3 and favoring a deviation of the flux toward the South Crater vents through S and C.

# Conclusions

Identifying warning signals announcing abrupt changes of hydrothermal activity of complex geological systems is crucial to improve risk evaluation of natural hazards. In this context, we have equipped the La Soufrière volcano in Guadeloupe by many geophysical sensors, including temperature probes in active vents and seismic array at the summit, both characterized by very short time responses, and muons telescopes located around the lava dome, offering invaluable capabilities to image density changes of this highly heterogeneous structure with a time resolution of few days[17]. By performing a permanent monitoring, we highlighted (i) a strong correlation with a delay of about 40 mn between the time fluctuations of temperature and seismic attributes over a particular 3-day period, associated to (ii) a particular hydrothermal activity of a rather small zone of about $10^4$ m$^3$ located few tens of meters below the surface and (iii) an abrupt increase of the muon flux, which means a strong decrease of density. To assess a quantitative description and interpretation of these results, we considered a Helmholtz-like resonator system. As a result, we estimate variations of 0.25% of steam mass fraction above a minimum of 11%: these fluctuations are followed by an abrupt and strong increase of 0.75%, which suggests a volume fraction of steam around 97%, assuming pressures less than 10 bars.

Observations of such short-term phenomena, described for the first time in the context of a hydrothermal systems, are not straightforward: in order to detect such abrupt changes, meteorological and anthropic noises and gaps in the measurements should be avoided, which remain very challenging in the environment of the La Soufrière volcano in Guadeloupe. Let us emphasize that cosmic muon radiography is less sensitive to the ambient conditions and appear as promising tools to detect short-term phenomena.

Further improvements will concern the simultaneous usage of several muon telescopes with high acceptance to improve the time resolution[18,19] to ultimately emit reliable warning messages. This will also provide a better 3D localization of the active hydrothermal focus and better identify the volcano sectors that may be prone to a high risk of short-term destabilization. Such quantitative results are not straightforward and remain very rare. Further improvements will concern extension of the seismic antennas and the monitoring of more fumaroles. The usage of other types of measurements, i.e. multi-gas and high time-resolution deformation, should also help to better constrain models and improve our understanding of this type of rapid phenomena which could produce dangerous events. Gravity monitoring could provide valuable information to constrain the mass budget and capture fast mass changes not detectable by the muon telescope. Monitoring of the electrical resistivity could also be a valuable method to improve the detection of liquid/vapor changes.

## Methods

**Blind independent processing protocol.**

In order to minimize biased interpretation during the data processing stages, we followed a blind independent protocol where the muon data and the seismic data used to localize the active zone shown in Fig. 3 have been processed independently by different members of the team: YLG and DG for both the seismic source localization and the RMS and time-frequency analysis; JM and MRC for the time variations of muon count in each line of sight of the telescope; DG for the log-periodicity analysis of the temperature time series of Fig. 2A. The merging of all data has subsequently been performed by DG to produce Fig. 2, 3 and 4. By this way, we believe that no interpretative bias has been introduced in our work.

**Cosmic muon radiography.**

Muon radiography relies on the same principles than classical X-ray medical radiography, and measures the attenuation of a beam of cosmic muons when crossing matter (here the lava dome) along trajectories $L_k$. Excepted for low energy muons or very dense matter (e.g. Pb, U, W), scattering is negligible and the trajectories $L_k$ may safely be considered as straight lines[29]. This makes the tomography problem much simpler and efficient for complex structures like a lava dome. The opacity $\varrho(\mathrm{kg.m^{-2}}) = \int_L \rho(\zeta) d\zeta$ of the geological structures is determined by comparing the muons flux $\Phi$ after crossing the target to the incident open sky flux, $\Phi_0$. The loss of energy of the muons along their trajectories through rock accounts for the standard bremsstrahlung, nuclear interactions, and e$^+$e$^-$ pair production physical processes, taken as,

$$\frac{-\partial E}{\partial \varrho}(\mathrm{MeV\,g^{-1}\,cm^2}) = a(E) + b(E) E,$$

where the functions a and b depend on the material properties [16]. The flux $\Phi$ of emerging muons is the part of the incident flux $\Phi_0$ of muons having an energy larger than $E_{\min}(\varrho)$, the minimum initial energy necessary to cross the opacity $\varrho$. As a rule of thumb, a muon loses about $2.2\,\mathrm{MeV}$ when crossing $1\,\mathrm{cm}$ of water.

The telescope used in the present study is equipped with three detection matrices of $10 \times 10$ pixels of $5 \times 5\,\mathrm{cm}^2$ formed by intersecting scintillator strips[30]. The two extreme matrices are parallel and separated by a distance $D = 1\,\mathrm{m}$. The matrices are synchronized on the same master clock signal with a timing resolution better than $1\,\mathrm{ns}$, and no shielding was used during these experiments since the telescope lies in a fault, a position which naturally filters low-energy particles forward scattering, the dominant background source[29]. Each detected muon has its trajectory $L_k$ determined by the pixels crossed by the particle in the front and rear matrices. For the present telescope, the combination of pixels that can be crossed define a set of $19 \times 19 = 361$ lines of sight forming the view-field covered by the telescope. The solid angle spanned by each line of sight depends on the distance between the front and rear matrices. The combination of this solid angle and the surface of the pixels contributing to a given line of sight defines the acceptance $(\mathrm{cm}^2\,\mathrm{sr})$ which quantifies the detection capacities of the telescope. In particular, the acceptance controls the space- and time-resolution through a feasibility formula[18,19]. Both the view-field and its acceptance are shown in Extended Data Fig. 1. In the present study, the telescope setup is: azimuth = 345°; inclination = 28.5°; XY$_{\mathrm{UTM}}$ = 20N 643115 1773938; $z = 1268\,\mathrm{m}$ above m.s.l. In order to increase the acceptance and improve the space- and time-resolution we merged lines of sight as indicated in Extended Data Fig. 1 where the red rectangle labeled 1 encloses the lines of sight crossing the active source area identified with the seismic data.

**Seismic measurements and data processing.**

The ambient seismic vibrations are sampled at 250 Hz frequency with an array of GS-11D (from Geo Space LP, USA) vertical geophones with a low-pass cut-off frequency of 3 Hz. The geophones are connected to Gantner A108 analog-to-digital 19 bits modules and the numerical data are transferred to the Gantner

QStation data concentrator through a RS485 serial data bus. A common GPS time base is used for both temperature and seismic data. Two groups of 8 geophones are disposed to form heptagonal antennas located East of the Tarissan crater and North of the Napoléon crater (NN antenna) and midway between Tarissan and South craters (POCS antenna). Each antenna has a diameter of about 30 m. Three other geophones are aligned along the fracture of the South crater. Because of their location on top of the lava dome, the geophones are very sensitive to noise induced by meteorological conditions and anthropic noise produced by the tourist walking around. The period considered in the present study was particularly favorable thanks to particularly quiet weather conditions.

Only the data of the POCS antenna have been used to compute the RMS and the dominant frequency time-series of Fig. 2B and 2C. The data of the South crater and of the NN antenna were discarded because of their too low signal-to-noise ratio. Both the RMS and the dominant frequency are computed for data segments of 20 s and after band-pass filtering the raw data in the 3-6 Hz frequency band where the coherency is maximum for all pairs of seismic time-series. Let us recall that spectral coherency is a correlation measure between the Fourier components of two signals. In the present study, the coherency is computed for packets of 30 Fourier transforms computed in 30 contiguous 20 s windows along the seismic time-series (see extended data Fig. 4).

The localization of the source of seismic noise is performed by back-propagating wavefronts delayed by the time-lags between the different geophones. The time lags are determined by searching the maximum value of the cross-correlation between all possible pairs of seismic time-series. In addition to the POCS antenna, 2 geophones of the NN and 2 of the South crater are used thanks to their acceptable signal-to-noise ratio. The cross-correlations are computed for 15 mn data segments (i.e. the duration of the data packets sent by the data loggers) and the time lags are averaged for all segments covering the 3-day period of data. Prior to cross-correlating, the raw data are band-pass filtered in the 3-25 Hz frequency band where significant spectral coherency is observed for most pairs of seismic time-series. By keeping the higher frequency as possible, we improve the time-resolution of the computed time-delays. Because of the lack of a seismic-velocity model for the lava dome, we adopt a constant velocity to back-propagate the delayed wavefronts. The algorithm then resumes to expanding hemispherical wavefronts and determine the loci of convergence. This is achieved by discretizing the lower half-space $5 \times 5 \times 5 \, m^3$ in cubic voxels and looking for those voxels where at least 8 wavefronts are simultaneously present. The source volume represented in Fig. 3 is the convex hull of all voxels satisfying this condition.

**Temperature measurements and processing.**

The temperature at the vents of the South crater is measured every second by Pt1000 probes inserted several tens of centimeters in the conduit of the vent in order to resist to the strong steam flux. The Pt1000 are connected with a 4-wires protocol to a Gantner A107 analog-to-digital 19 bits module. The numerical data issued by the module are transferred to a data concentrator Gantner QStation through a RS485 serial data bus with a length of about 100 m. The temperature time-series are continuously recorded since May 2015.

Defining a cycle as the time period separating two successive temperature relative minima, 10 such cycles may be identified in the curve of Fig. 2A. Extended Data Table 1 summarize the characteristics of each cycle. Extended Data Fig. 2A shows that the duration of the successive cycles clearly increases from 3 hours (cycle 1) to 8 hours (cycle 9). Dividing a cycle into a fall period and a rise period, we observe that the fall period remains almost constant with an average value of about 2 hours (Extended Data Fig. 2B) while the rise period progressively increases from 2 hours to 6 hours when progressing along the series of cycles (Extended Data Fig. 2C). Consequently, the positive trend visible in the total cycle duration (Extended Data Fig. 2A) is mainly due to the positive trend of the rise period (Extended Data Fig. 2C). A positive trend of temperature is superimposed on the cycle sequence with an average temperature rise of 0.1 C between successive temperature minima (Extended Data Fig. 2D). The temperature decrease that occurs during the fall period of the cycles steadily augments from 0.15 C to 0.4 C (Extended Data Fig. 2E). Meanwhile, the temperature increase occurring during the rise period of the cycles also augments from 0.15 C to 0.65 C (Extended Data Fig. 2F). To summarize, all cycles attributes display a clear positive trend excepted for the fall period which remains almost constant for all cycles (Extended Data Fig. 2B).

The pattern observed in the temperature oscillations of increasing period resembles to the anti-bubble situation seen in financial time series where and corresponding to a reverse log-periodic sequence[31]. Owing to this reverse-time situation, the log-periodicity of the temperature cycles can be tested by checking if the start times of the cycles are arranged as,

$$t_n - t_c = \tau \times \lambda^n$$

where *n* is the cycle index increasing with time, $\tau$ is the time unit, $\lambda$ is the scaling ratio and $t_c$ is a critical time marking the onset of some critical transition. Remarking that,

$$\lambda = \frac{t_{n+2} - t_{n+1}}{t_{n+1} - t_n},$$

and using the successive start times of the cycles (see Extended Data Table 1) we find an average $\lambda = 1.3$. The critical time can be derived through a Shanks transformation[32,33],

$$t_c = \frac{t_{n-1} \times t_{n+1} - t_n^2}{t_{n-1} + t_{n+1} - 2 t_n}.$$

In order to make the critical time estimate more robust, we use the parameters of the linear fit to the cycle durations (see Extended Data Fig. 2A) instead of the individual values $t_n$. We find $t_c \approx -15\,\text{h}$. This indicates that the sequence of temperature cycles of increasing period critically converges backward in time to date roughly half a day before the onset of the observed oscillations.

**Data availability.**

The research data relevant to this Letter are available upon request to the corresponding author DG (gibert@univ-rennes1.fr).

## Acknowledgements

This study is part of the DIAPHANE project ANR-14-CE 04-0001. DEM data and meteorological data are provided by the OVSG. Anonymous Referees made very constructive comments.

## Author contributions

YLG and DG performed the seismic data analysis and modelings. MRC and JM performed the muon data processing. YLG, DG and JBA contributed to the temperature and seismic data interpretation. DG proceeded to the integration of the results of all data analysis. BK constructed and implemented seismic sensors and electronic devices. BC was in charge of the online distributed DAQ-processing software of the muon telescopes, based on the work of T. Descombes from LPSC (UMR 5821, CNRS-IN2P3, Université Grenoble Alpes). All authors participated to the implementation of the arrays of sensors on the field. DG, JM, JCI and MRC ensured the maintenance of equipments on the field. All authors reviewed the manuscript. DG is PI of the DIAPHANE project. One of the author, JM, wants to thank C. Dufour for her long-standing contribution to the project.

**Author information**

The authors may be joined at the following e.mails: yves.legonidec@univ-rennes1.fr; rosas@ipgp.fr; bremond@univ-rennes1.fr; carlus@ipnl.in2p3.fr; ianigro@ipnl.in2p3.fr; bruno.kergosien@univ-rennes1.fr; marteau@ipnl.in2p3.fr; dominique.gibert@univ-rennes1.fr.

**Competing Interests:** The authors declare no competing interests.

# SUPPLEMENTARY DATA

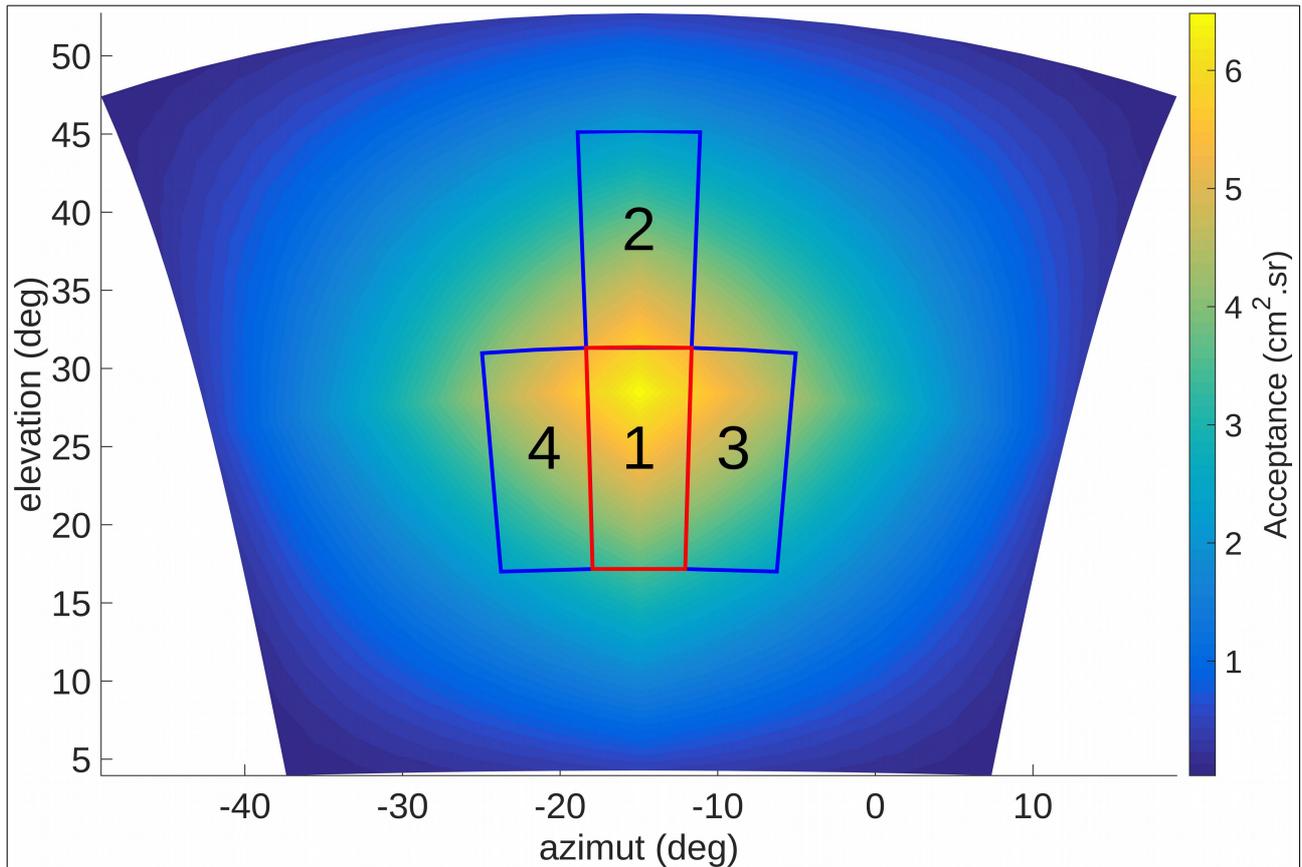

**Extended Data Fig. 1** | Telescope acceptance function and merged lines of sight. View field covered by the $19^2 = 361$ lines of sight of the cosmic muon telescope. The red rectangle encompass the lines of sight covering the seismic source zone of Fig. 3. These lines are merged to obtain a joined acceptance of 54.3 $cm^2 sr$. The time variations of the muons count in this area correspond to the red curve in Fig. 4. The blue rectangles located above and on both sides of area 1 have muons counts shown as blue curves in Fig. 4. The joined acceptances of areas 2, 3 and 4 are respectively: 34.8 $cm^2 sr$, 40.3 $cm^2 sr$ and 45.8 $cm^2 sr$. The maximum acceptance of the axial line of sight of the telescope equals 6.5 $cm^2 sr$.

| Cycle # | Start date (March 2017 UTC) | End date (March 2017 UTC) | Date of maximum (March 2017 UTC) | Duration (h) | Rise time (h) | Fall time (h) | Rise temperature (C) | Fall temperature (C) |
|---|---|---|---|---|---|---|---|---|
| 1 | 28 17:33 | 28 20:32 | 28 19:22 | 2.98 | 1.82 | 1.17 | 0.12 | 0.13 |
| 2 | 28 20:32 | 29 00:00 | 28 22:23 | 3.47 | 1.85 | 1.62 | 0.13 | 0.14 |
| 3 | 29 00:00 | 29 04:20 | 29 02:59 | 4.33 | 2.98 | 1.35 | 0.25 | 0.16 |
| 4 | 29 04:20 | 29 09:05 | 29 07:23 | 4.75 | 3.05 | 1.70 | 0.20 | 0.20 |
| 5 | 29 09:05 | 29 14:25 | 29 12:51 | 5.33 | 3.77 | 1.57 | 0.36 | 0.12 |
| 6 | 29 14:25 | 29 19:16 | 29 17:35 | 4.85 | 3.17 | 1.68 | 0.15 | 0.19 |
| 7 | 29 19:16 | 30 03:20 | 30 00:05 | 8.07 | 4.82 | 3.25 | 0.32 | 0.34 |
| 8 | 30 03:20 | 30 12:25 | 30 10:17 | 9.08 | 6.95 | 2.13 | 0.41 | 0.26 |
| 9 | 30 12:25 | 30 20:38 | 30 18:39 | 8.22 | 6.23 | 1.98 | 0.47 | 0.39 |
| 10 | 30 20:38 | 31 05:35 | 31 02:40 | 8.95 | 6.03 | 2.92 | 0.65 | 0.06 |

**Extended Data Table 1** | Data of the temperature cycles observed in the time series of Fig. 2A.

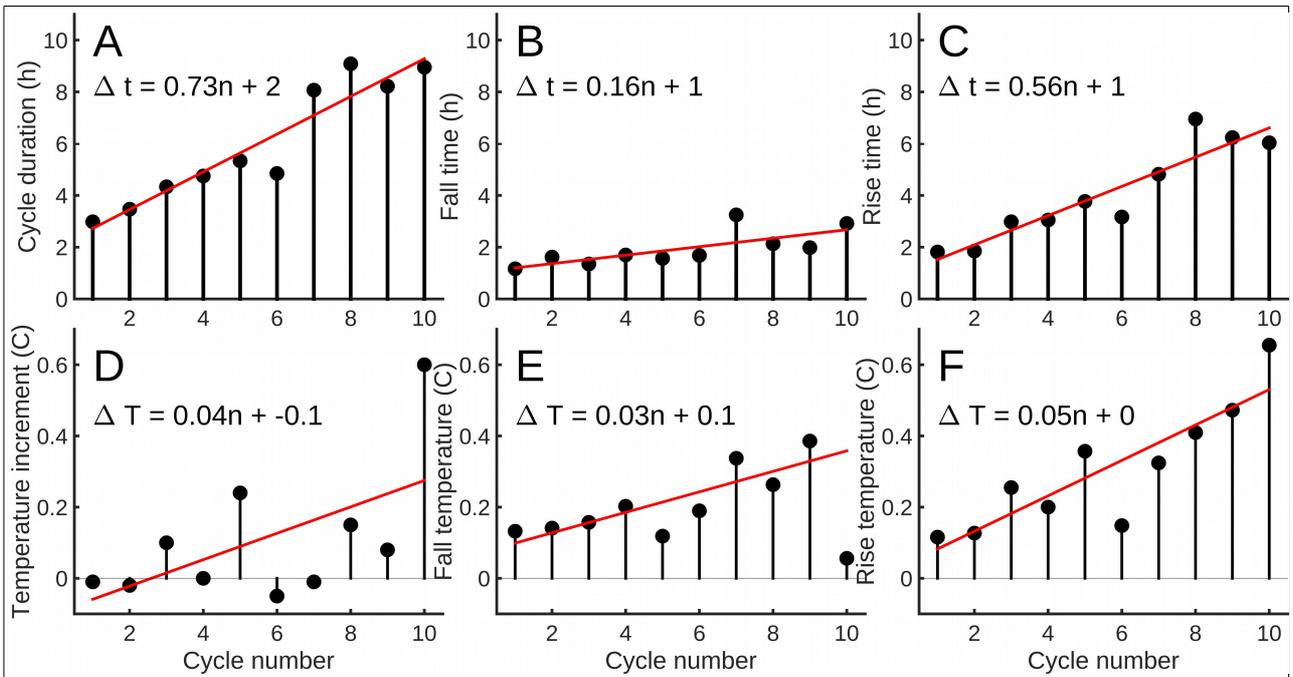

**Extended Data Fig. 2** | Main characteristics of the temperature cycles observed in the time series of Fig. 2A. A cycle is defined as period separating two successive minima in the temperature time series of Fig. 2A. By this way, 10 cycles are identified (see Extended Data Table 1 for data values). A: Cycle duration (in hours) defined as the interval between successive temperature minima. B: Temperature fall time (in hours) separating a temperature maximum from the next minimum. C: Temperature rise time (in hours) separating a temperature minimum from the next maximum. D: Increase of temperature between successive minima. E: Temperature decrease between a maximum and the next minimum. F: temperature increase between a minimum and the next maximum. Red straight lines represent the best first-order polynomial fit. Polynomial formula is indicated in each plot. The fall temperature (E) observed during the decreasing phase of the cycles displays a positive trend which, when added with the global positive trend (D), compensates the trend of the rise temperature (F). This indicates that, when corrected for the global trend, a cycle returns to its starting temperature. The temperature time series of Fig. 2A is them the superimposition of a trend and the sequence of oscillations. The temperature trend may be due to a constant steam supply causing a progressive increase of the pressure in the source zone of Fig. 3.

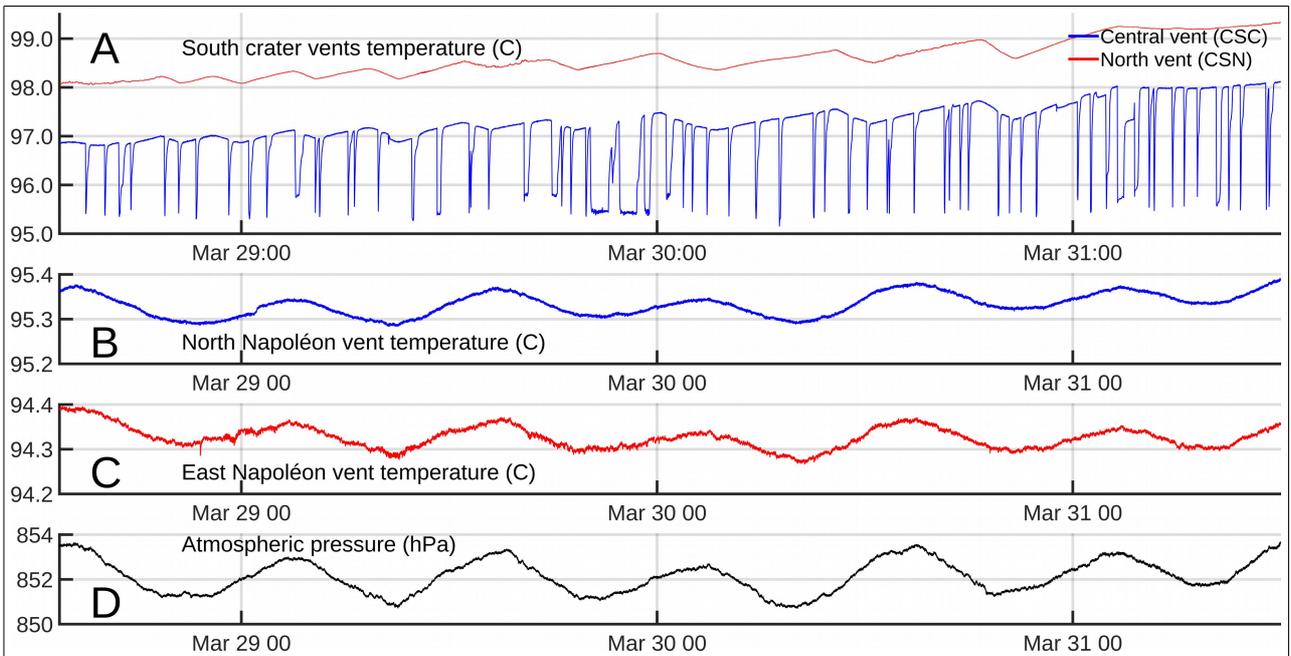

**Extended Data Fig. 3** | Atmospheric pressure and Temperature time series in vents. See Fig. 1 for location of the vents. A: Temperature time-variations in the Central and North vents of the South crater. The temperature oscillations observed in the North vent time-series are retrieved in the upper envelope of the Central vent time-series. B: Temperature time-variations in the North Napoléon vent. C: Temperature time-variations in the East Napoléon vent. D: Time-variations of the atmospheric pressure measured at the data concentrator location. Both the North and East Napoléon vents have a low-pressure flux and do not display the remarkable oscillations visible in the vents of the South crater. Instead, the small temperature variations of the Napoléon vents are in phase with the fluctuations of atmospheric pressure.

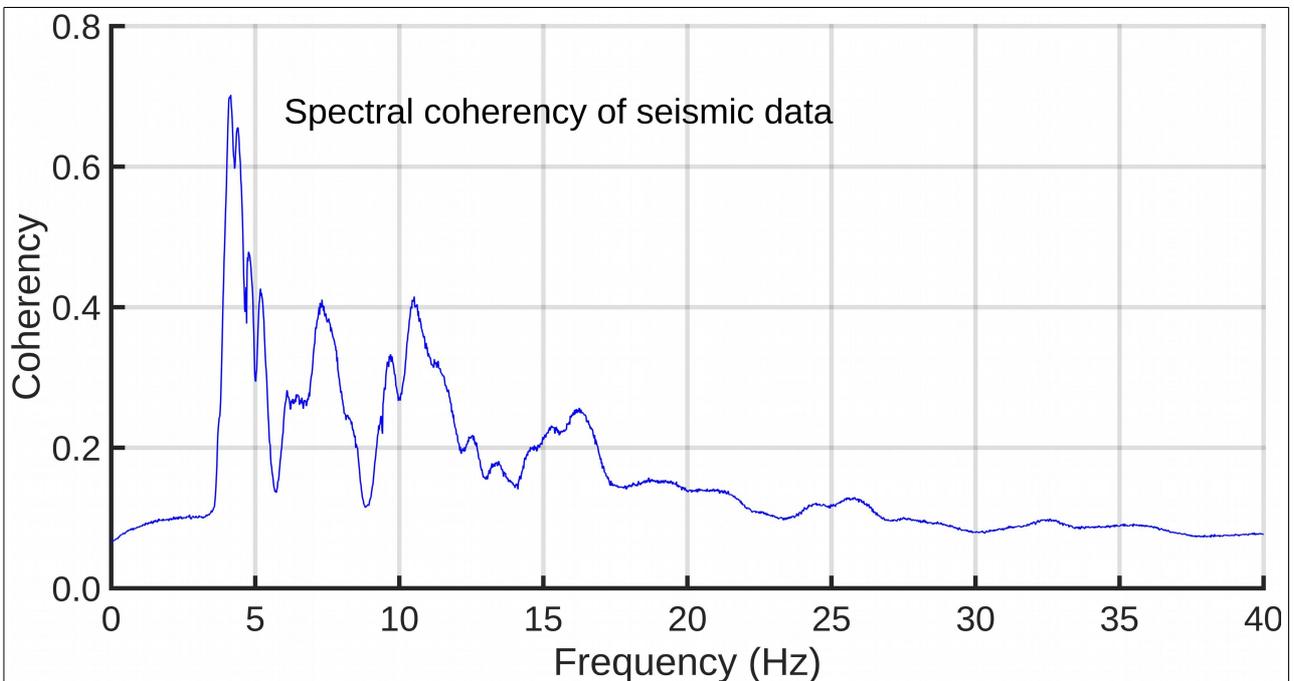

**Extended Data Fig. 4** | Spectral coherency of the seismic noise data. The spectral coherency is computed for all seismic data of the POCS antenna according to the method described in Bendat and Piersol[14]. Harmonic spectral lobes are clearly visible.

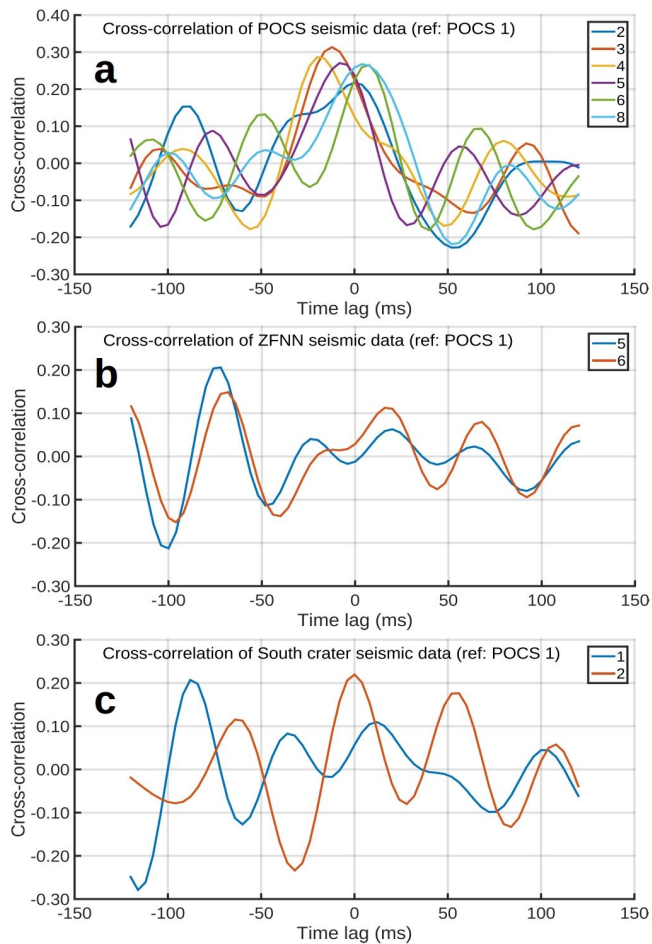

**Extended Data Fig. 5** | Cross-correlations of the seismic data. The graphs represent the cross-correlation functions of the geophone data used to localize the source of seismic noise (Fig. 3). All cross-correlations have been computed with respect to geophone 1 of the POCS antenna (Fig. 1). The data are band-pass filtered in the 3-25 Hz frequency band where the spectral coherency is significant (Extended Data Fig. 4). The geophones that are not been used in this study are discarded because of either an electrical problem (POCS #7 and NN #3, 8] or a too low signal-to-noise ratio possibly caused by a poor mechanical coupling with the unconsolidated granular soil and insignificant cross-correlation (SC #3, NN #1, 2, 4, 7).

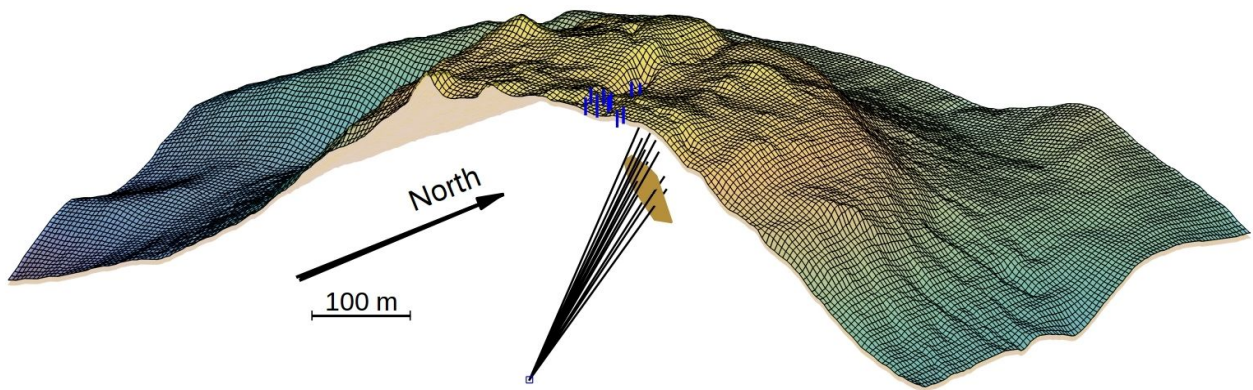

**Extended Data Fig. 6** | Location of the active hydrothermal spot in the La Soufrière lava dome. See Fig. 3 for other details.